\documentclass[conference]{IEEEtran}
\IEEEoverridecommandlockouts
\usepackage{cite}
\usepackage{amsmath,amssymb,amsfonts}
\usepackage{algorithmic}
\usepackage{graphicx}
\usepackage{textcomp}
\usepackage{xcolor}

\usepackage{booktabs}
\usepackage{multirow}
\usepackage{hyperref}

\usepackage{subcaption}

\usepackage[normalem]{ulem}
\usepackage{etoolbox}

\newif\ifcmgfinal\cmgfinalfalse

\newcommand{\cmgmark}{\textsuperscript{\textcolor{blue}{\tiny\textsf{CMG}}}}

\ifcmgfinal
  \newcommand{\cmg}[1]{}
  
  \newcommand{\cmgremove}[1]{}
  
\else
  \newcommand{\cmg}[1]{\textcolor{blue}{\scriptsize \slshape {[CMG: #1]}}}
  
  \newcommand{\cmgremove}[1]{\textcolor{blue}{\sout{#1}}\cmgmark}
  
\fi
\robustify\uline
\robustify\sout

\def\BibTeX{{\rm B\kern-.05em{\sc i\kern-.025em b}\kern-.08em
    T\kern-.1667em\lower.7ex\hbox{E}\kern-.125emX}}
\begin{document}

\newif\ifblind

\ifblind
\else
 \title{How well do routinely collected demographic and clinical variables aid point-of-care lung ultrasound TB classification
}
\fi


\newcommand{\orcidauthor}[2]{#1\hspace*{0.6mm}%
  \href{https://orcid.org/#2}{\includegraphics[width=0.9em]{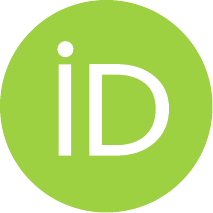}}}

\newcommand{\affilsun}{%
  \textit{Department of Electrical and Electronic Engineering},
  \textit{University of Stellenbosch}, Stellenbosch, South Africa}
\newcommand{\affilepfl}{%
  \textit{Laboratory for intelligent Global Health and Humanitarian Response
  Technologies (LiGHT)}\\
  \textit{Swiss Federal Institute of Technology (EPFL)}, Lausanne, Switzerland}
\newcommand{\affilctr}{%
  \textit{South African Medical Research Council Centre for Tuberculosis
  Research (CTR)}\\
  \textit{Division of Molecular Biology and Human Genetics, Faculty of Medicine
  and Health Sciences}\\
  \textit{Stellenbosch University}, Cape Town, South Africa}
\newcommand{\affilcnhu}{%
  \textit{National Teaching Hospital for Tuberculosis and Pulmonary Diseases
  (CNHU-PPC)}, Cotonou, Benin}

\newcommand{\emailaddr}[1]{{\footnotesize\normalfont\texttt{#1}}}

\ifblind
  \author{\IEEEauthorblockN{Anonymous Author(s)}
  \IEEEauthorblockA{Affiliation withheld for double-blind review}}
\else
    \author{
    \IEEEauthorblockN{
      \orcidauthor{Joshua M. Jansen van Vüren}{0000-0002-3406-4788}\IEEEauthorrefmark{1},
      \orcidauthor{Christiaan M. Geldenhuys}{0000-0003-0691-0235}\IEEEauthorrefmark{1},
      \orcidauthor{Devendra S. Parihar}{0000-0001-6404-6428}\IEEEauthorrefmark{1},
      \orcidauthor{Véronique Suttels}{0000-0002-5261-0439}\IEEEauthorrefmark{2},\\
      \orcidauthor{Trevor Brokowski}{0000-0002-7547-7749}\IEEEauthorrefmark{2},
      \orcidauthor{Ablo P. Wachinou}{0000-0002-1024-9665}\IEEEauthorrefmark{4},
      \orcidauthor{Mary-Anne Hartley}{0000-0002-8826-3870}\IEEEauthorrefmark{2},
      \orcidauthor{Rensu P. Theart}{0000-0003-0508-1690}\IEEEauthorrefmark{1},\\
      \orcidauthor{Grant Theron}{0000-0002-9216-2415}\IEEEauthorrefmark{3},
      \orcidauthor{Thomas R. Niesler}{0000-0002-7341-1017}\IEEEauthorrefmark{1}}%
    \IEEEauthorblockA{\IEEEauthorrefmark{1}\affilsun}%
    \IEEEauthorblockA{\IEEEauthorrefmark{2}\affilepfl}%
    \IEEEauthorblockA{\IEEEauthorrefmark{4}\affilcnhu}%
    \IEEEauthorblockA{\IEEEauthorrefmark{3}\affilctr}%
    \IEEEauthorblockA{%
      {\footnotesize\normalfont Corresponding authors:}\hspace*{0.35em}%
      \emailaddr{jjvanvueren@sun.ac.za}, \emailaddr{trn@sun.ac.za}}%
    }
\fi

\maketitle

\begin{abstract}
We consider the fusion of lung ultrasound images with routinely-collected clinical and demographic data for the purpose of automated tuberculosis (TB) screening using deep-learning.
Such deep-learning based screening tools for TB could meaningfully support the health care system in Africa, where the burden of disease is severe and resources are constrained.
Beginning with an established ResNet baseline for classification of lung ultrasound images, which achieves an area under the receiver operating characteristic (AUROC) curve of 0.91 [0.86,0.96] (95\% CI), we consider the incorporation of the clinical and demographic data using three fusion approaches.
We find that a simple average-based fusion of the output scores of separately-trained image and clinical data classifiers consistently matches or outperforms a more complex approach where the data is fused earlier and a combined classifier is trained.
Fusing the image and the clinical classifiers in this way leads to a classifier with an overall AUROC of 0.95 [0.91,0.99] (specificity of 0.76 at sensitivity 0.93) which is an improvement of 4\% absolute over the image-only baseline.
We also find that greedy feature selection can be used to reduce the number of clinical and demographic inputs without sacrificing classification performance.
Finally, when we differentiate between clinical and demographic data that are self-reported, that require some basic measurement or calculation, and that require a point-of-care (POC) test, we find the inclusion of the POC tests included in this study to be of minimal benefit to classification performance.
We conclude that the incorporation of routinely-collected clinical and demographic data is a promising way to improve the performance of lung ultrasound based automatic classification.
\end{abstract}

\begin{IEEEkeywords}
ResNet, tuberculosis, screening, clinical and demographic data, fusion, lung ultrasound
\end{IEEEkeywords}

\section{Introduction}
In 2024, TB was once again the world's leading cause of death from a single infectious agent~\cite{tb_report}.
Accessible, scalable and cost-effective screening and triage tools for TB therefore remain a priority, especially in Africa, a region accounting for 25\% of the 10.7 million people who fell ill with the disease that year.
The application of basic clinical data or results from laboratory-based tests as a method to identify people with TB has been reported in the literature, including the application of various machine learning approaches~\cite{Bartl2025-jo,Orjuela-Canon2025-eh,botha2018,Ghermi2024-zv,Parihar2025}.

Recently, handheld point-of-care ultrasound (POCUS) devices have become readily available in many clinical settings~\cite{Lee2020}.
Their use in echocardiography and emergency medicine, for example, is well established.
POCUS has been employed for use in identifying haemothorax~\cite{Brooks44} and pneumothorax~\cite{Zhang2006} in people with thoracic trauma.
Ultrasonography of the lung relies on the identification of specific patterns to enable similar insights, for example the identification of comet-tails, lung sliding, and anechoic areas~\cite{Staub2018,Lichtenstein1995,Lichtenstein1998}.

It is possible, therefore, that deep-learning image-based analysis  may be able to accomplish these tasks automatically, as they are strong pattern recognition systems.
Indeed, some deep-learning based approaches developed during the COVID-19 pandemic have achieved high performance~\cite{jimaging8030065,KHAN2023106994}.
For example, in a cohort of  $n=\ $48 patients admitted to ICU with COVID-19 like symptoms, and for whom 692 lung ultrasound (LUS) scans were captured, an AI-based tool (ExoLungAI) was able to identify A-lines with 0.91 sensitivity and 0.81 specificity and B-lines with 0.84 sensitivity and 0.86 specificity~\cite{diagnostics14222526}.
Similarly, when applied to $n=\ $28 COVID-19 positive patients, a deep-learning based model was shown to have high agreement with a human expert in identifying lung pathologies such as B-lines, consolidations and pleural effusion.
This model was also shown to effectively augment the examination process by reducing the average time required from 12.6 to 5.3 minutes~\cite{ijtm2010002}.
Finally, a sensitivity of 0.88 and a specificity of 0.94  was achieved when employing VGG to classify frames from LUS scans of 1204 COVID-19 patients against similar data from 2030 people who were either healthy or had bacterial pneumonia~\cite{app11020672}. 
However, when this same model is applied to a group of $n=\ $66 patients, 33 of whom are positive for COVID-19, a sensitivity of 0.62 and specificity of 0.7 was achieved, showcasing the challenge of developing such systems which are robust to applications in unseen settings.
After fine-tuning, a sensitivity of 0.9 and specificity of 0.55 ~\cite{covid_blues} was achieved.

LUS shows promise as a practical screening modality, as the handheld devices are both mobile and more cost-effective than classical ultrasound devices~\cite{BARIBEAU20203139}.
However, their effectiveness for the detection of pulmonary TB is to date unclear, and further research is needed~\cite{ptb_diagnostic_accuracy_review}.
To our knowledge, only two studies have considered the application of deep-learning to predict TB from LUS.
The first employed a CNN-LSTM with a cohort of $n=\ $710 patients to classify bacteriologically or clinically confirmed TB from 11455 videos, achieving a sensitivity of 0.92 and specificity of 0.73 (AUROC=0.89) in a test set of 156 people~\cite{swinkels2025prototype}.
The second, a prospective study employing the same data used in this work, found a ResNet-based TB classifier trained on still images captured during a LUS exam to afford an AUROC of 0.91 [0.89,0.93] (95\% CI) in a cohort of $n=\ $504 patients, with a test set of $n=\ $101~\cite{ultrai2025escmid}.

In this work we make the following contributions.
\begin{itemize}
    \item We show that ResNet-based models afford similar performance when trained on a composite reference standard of bacteriologically and clinically diagnosed TB to previous work using only bacteriologically confirmed TB.
    \item We demonstrate that classifiers based only on clinical and demographic data yield reasonable performance.
    \item We demonstrate that small but consistent improvements are seen when clinical and LUS classifiers are fused.
    \item We show that the simple averaging of the output probabilities from the two classifiers outperforms more complex ways to incorporate the clinical feature vector directly into the LUS classifier.
\end{itemize}

\section{Data}

The dataset we employ comprises $n=\ $502 patients presenting with respiratory symptoms to the National Teaching Hospital for Tuberculosis and Pulmonary Diseases (CNHU-PPC) in Benin~\cite{ultrai2025escmid,suttels2022imagebanks}.
POCUS exams were conducted using a standardised 14-point sliding protocol and images were captured.
TB status was determined using a sputum-based Xpert MTB/RIF Ultra~\cite{Dorman2017} test.
Additionally, participants classified with clinically diagnosed TB (based on clinical, radiological, and outcome data) are treated as positive in this cohort.
This inclusion of clinically diagnosed TB differentiates the work presented here from previous results on the same dataset, in which only Ultra results were used to determine TB.
Data is partitioned into five folds for cross-validation, with an explicit fixed test set.
The total and per-fold counts for the number of participants, those with confirmed TB, and number of LUS scans are presented in Table~\ref{tab:cv_stats}.

\begin{table}[]
    \centering
    \caption{Cross validation fold statistics. Num scans: mean [min,max].}
    \label{tab:cv_stats}
    \begin{tabular}{c c c c}
        \toprule
        Fold & TB Pos   & \# Patients & Num scans [min,max] \\
        \midrule
        1    & 32       & 81  & 27 [13, 31] \\
        2    & 36       & 80  & 27 [18, 31] \\
        3    & 37       & 80  & 26 [10, 33] \\
        4    & 40       & 80  & 27 [15, 33] \\
        5    & 36       & 80  & 27 [14, 31] \\
        \midrule
        Test & 43       & 101 & 27 [19, 33] \\
        \midrule
        Total& 224      & 502 & 27 [10, 33] \\
        \bottomrule
    \end{tabular}
\end{table}

For each participant, a series of additional clinical, demographic and POC tests were collected (Table~\ref{tab:clinical_features}).
All features which were available for fewer than 95\% of patients for any fold were not considered.
Median imputation was used to provide the remaining missing values, which is suitable due to skewed distributions present in those columns for which missing values occur.
No imputation was performed for one-hot encoded values.
The variables with the most missing values were SpO$_2$ and whether traditional medicine was sought, each missing for only 3.1\% of the total dataset.
As we would like to determine which of the features listed in Table~\ref{tab:clinical_features} aid a deep-learning based LUS system to triage for TB, we partition them into subsets based on their ease of access.
The first,  class A, includes variables which are easiest to acquire and are self-reported on a questionnaire.
This class therefore is anamnestic in nature and subject to the patient's ability to accurately recall.
The second, class B, includes variables which require some basic measurement, or the intervention of a minimally trained healthcare worker.
Finally, class C includes any value which requires a POC test, and represents the least accessible of the features we consider.
It is important to note that for this study only two POC tests were considered (HIV and malaria) and therefore the conclusions drawn are only with respect to these two tests, and not to POC tests in general.
The division into these classes is relevant because easy-to-acquire variables minimise the barrier to downstream molecular testing and true TB verification.
Therefore, the identification of such features that also improve a deep-learning LUS screening tool reduce the burden on and improve the throughput of the healthcare system.

\begin{table}[h!]
    \caption{Clinical, demographic and POC values collected for each of the participants in the Benin dataset~\cite{suttels2022imagebanks}. Variables are partitioned into one of three classes (A: 28,B:13,C:3) in order of their ease-of-acquisition, A being the most accessible and C the least. }
    \label{tab:clinical_features}
    \centering
    \setlength{\tabcolsep}{2pt}
    \begin{tabular}{l l l}
        \toprule
        Class & Kind & Variable \\
        \midrule
        \multirow{28}{*}{A}
        & Numeric   & Age \\
        & Binary    & Asthenia \\
        & Binary    & Anorexia \\
        &One-hot    & Chief complaint (\texttt{\scriptsize cough, fever, etc.}) \\
        & Binary    & Cardiopathy \\
        & Binary    & Chronic lung disease \\
        & Binary    & Cough \\
        & Numeric   & Cough days \\
        & Binary    & COVID vaccinated \\
        & Binary    & Diabetes \\
        & Binary    & Dyspnoea \\
        & Binary    & Chest pain \\
        & Binary    & Fever \\
        & Numeric   & Fever days \\
        & Binary    & GI symptoms \\
        & Binary    & Haemoptysis \\
        &One-hot    & HIV status known (\texttt{\scriptsize no, yes - ART, yes - No ART}) \\
        & Binary    & Hypertension \\
        & Binary    & Night sweats \\
        & Numeric   & Ordinal general state (\texttt{\scriptsize 1:no change,$\dots$, 5:bed-bound}) \\
        &One-hot    & Previous TB (\texttt{\scriptsize no,yes - active, yes - resolved}) \\
        &One-hot    & Recent antibiotics (\texttt{\scriptsize no, yes: ongoing, yes: stopped}) \\
        &One-hot    & Sex (\texttt{\scriptsize female, male}) \\
        &One-hot    & Smoker (\texttt{\scriptsize never, quit > 6 months, passive, active}) \\
        & Binary    & Sore throat \\
        & Binary    & Taste loss \\
        & Binary    & Traditional medicine sought \\
        & Binary    & Weight loss \\
        \midrule
        \multirow{13}{*}{B}
        & Numeric   & BMI \\
        & Numeric   & Blood oxygen (SpO$_2$) \\
        & Binary    & Cervical or axillary adenopathy \\
        & Numeric   & Diastolic blood pressure \\
        & Numeric   & Glasgow Coma Scale~\cite{TEASDALE197481} \\
        & Numeric   & Height \\
        & Numeric   & Heart rate \\
        & Binary    & Pitting oedema \\
        & Numeric   & qSOFA~\cite{singer_qsofa} \\
        & Numeric   & Respiratory rate \\
        & Numeric   & Systolic blood pressure \\
        & Numeric   & Temperature \\
        & Numeric   & Weight \\
        \midrule
        \multirow{3}{*}{C}
        & Binary    & HIV \\
        & Binary    & Malaria RDT \\
        & Binary    & Malaria RDT done/not done \\
        \bottomrule
    \end{tabular}
\end{table}

\section{Experimental setup}

All experiments are conducted within a five-fold cross-validation scheme, where training occurs using four of the five folds and the fifth is held-out for development.
Hyperparameters are chosen based on the average development set loss across the five folds, fixed, and the development set re-introduced for final training on all five folds.
This final classifier is then applied to the unseen test set.
For all models, the AUROC is reported for the development and test sets.
A bootstrapped confidence interval of the AUROC is constructed for development sets by resampling 2000 times over the five sets, each time recomputing the average development set AUROC utilising the Mann-Whitney formulation.
From this empirical distribution, the values for the 2.5th and 97.5th percentile are determined~\cite{efron1994introduction}.
The reported confidence interval on the test set uses the DeLong-based method~\cite{delong_1988}.

\subsection{Clinical data classifier}

We evaluate both a simple logistic regression (LR) model as well as a 2 or 3 layer multi-layer perceptron (MLP) which receive as input the encoded clinical data as a feature vector.
Quantitative features are used directly, while binary or one-hot encoding is used for categorical variables, as denoted in Table~\ref{tab:clinical_features}.
Mean and variance normalisation, based on the training set, is applied to all feature vectors.
Both models are optimised over the hyperparameters listed in Table~\ref{tab:hyperparameters}.
After subsequent feature selection, hyperparameters are re-optimised over the same ranges.
The hyperparameters for the clinical-based models are not jointly optimised during feature selection due to the associated computational complexity.

\begin{table}[]
    \centering
    \setlength{\tabcolsep}{3pt}
    \caption{Model and optimisation hyperparameters for clinical (LR,MLP) and LUS TB-classification models. Parameters are presented for the top performing configuration. For the clinical classifiers these values are for the second stage sweep after feature selection has occurred.}
    \label{tab:hyperparameters}
    \begin{tabular}{llcc}
        \toprule
        Model & Hyperparameter      & Range & Optimal \\
        \midrule
            & Learning rate         & [0.1, 0.01, 0.001]                                    & 0.001 \\
        LR  & Batch size             & [8, 16, 32]                                          & 16 \\
            & Epochs                & [1, $\dots$, 128]                                     & 100 \\
        \midrule
                                & Learning rate                 & [0.01, 0.001, 0.0001]     & 0.001 \\
        \multirow{4}{*}{MLP}    & Batch size                    & [8, 16, 32]               & 16 \\
                                & \multirow{2}{*}{Hidden size}  & [[16],[32], [64, 32],     & \multirow{2}{*}{[16]} \\
                                &                               & [32, 32], [32, 64]] \\
                                & Dropout                       & [0.0, 0.3]                & 0.0 \\
                                & Epochs                & [1, $\dots$, 128]                 & 31 \\
        \midrule
        \multirow{6}{*}{LUS}
                                & Learning rate                 & [0.001, 0.0001]           & 0.0001    \\
                                & Batch size                    & [16, 32, 64]              & 16        \\
                               & Hidden size                   & [ 128, 256, 512]           & 128       \\
                                & Freeze feature extractor      & True, False               & False     \\
                                & Feature extractor             & [ResNet-18, ResNet-50]    & ResNet-18 \\
                                & Epochs                        & [1, $\dots$, 20]          & 6 \\
        \bottomrule
    \end{tabular}
\end{table}

\subsection{Feature selection}

For each of the clinical data classifiers and combinations (A, B, C, AB, AC, BC, ABC) we conduct sequential feature selection (SFS)~\cite{FERRI1994403}.
This algorithm begins with a single feature and greedily selects a subset of features based on those which achieve the lowest average development set loss.
Typically, the best development set performance is achieved for a subset of the features, after which additional features degrade performance.

\subsection{Lung ultrasound classifier}

We employ the LUS classification architecture presented in~\cite{ultrai2025escmid}.
This architecture represents a multi-instance learning approach where the $N'$ scans from a single patient are denoted as the unordered set $\boldsymbol{X}$ ($\mathbb{R}^{(N' \times H \times W \times C)}$).
Each scan has input sizes of $H \times W$, which is always resized to 224$\ \times\ $224, and $C =\ $1 because input scans are greyscale.

Each scan in the set is passed through an augmentation layer which imposes brightness and contrast jitter and random translations and rotations~\cite{Shorten2019}.
Subsequently, mean and variance normalisation is applied and scans are fed separately through a feature extractor, which is either a ResNet-18 or a ResNet-50~\cite{He_2016_CVPR,Russakovsky2015}.
The output vectors from these ResNets are either 512 (for ResNet-18) or 2048 (for ResNet-50) dimensional, and are projected to $h_d=\ $128, 256, or 512 such that the TB classification layers maintain the same number of trainable parameters during hyperparameter optimisation.

An embedding, unique to each of the 14 anatomical scan locations, with the same dimensionality ($h_d$) is then added to the feature vectors from each respective site.
These $N'$ feature vectors pass through two linear layers.
The output of the first increases the dimensionality to four times that of the input (4$\ \cdot\ $$h_d$) and then applies a ReLU activation.
The second layer decreases the dimensionality of the vector back to $h_d$.
Subsequently, an additive style attention is applied, which aims to select the scans which are most useful for TB classification.
This converts the set of $N'$ features to a single feature vector~\cite{bahdanau2014neural}.
Finally, this single feature vector is passed through a two layer classifier head, which again increases and then decreases the dimensionality of the vectors, this time by a factor of two.
The output of the classifier is a scalar sigmoid approximating the posterior probability of TB.

This classifier is optimised over the hyperparameters listed in Table~\ref{tab:hyperparameters} in a grid-fashion.

\subsection{Fusion of classifiers}
\label{subsec:exp_setup_fused_classifier}

\begin{figure}
    \centering
    \includegraphics[width=0.8\linewidth]{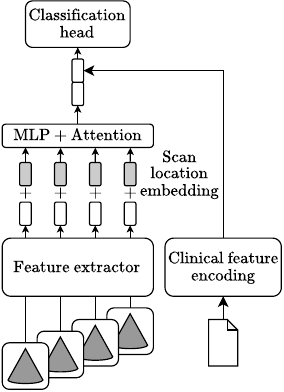}
    \caption{Fusion of clinical data into ultrasound classifier through the concatenation of the encoded feature vector before the final classification layers (classification head).}
    \label{fig:fusion_classifier}
\end{figure}

\begin{figure}
    \centering
    \includegraphics[width=0.8\linewidth]{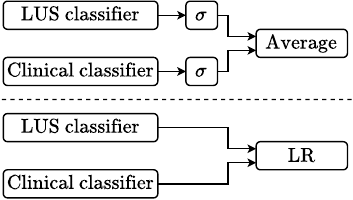}
    \caption{Alternative strategies for classifier fusion using either (top) a simple average of the posterior TB probabilities produced by the clinical and LUS-based classifiers or (bottom) LR trained directly on the model logits. The sigmoid function is denoted as $\sigma$.}
    \label{fig:fusion_alternative_exp_setup}
\end{figure}

We consider the fusion of the LUS and clinical data classifiers using three approaches.
First, the encoded feature vector of dimensionality ($h_f$) is concatenated with the feature vector obtained after the attention pooling layer, as illustrated in Figure~\ref{fig:fusion_classifier}.
The result of this concatenation is then fed to the classification head as previously described.
Second, the average of the two output probabilities produced by the LUS and the clinical classifiers is considered.
Finally, LR is trained on the output logits of the two classifiers (before the sigmoid activations).
The final two alternative configurations are illustrated in Figure~\ref{fig:fusion_alternative_exp_setup}.
This model is trained on the output scores produced for the development data (i.e. out-of-fold) by each classifier, again within a cross-validation scheme.
Out-of-fold scores are utilised as classifiers often produce overconfident scores on the training data which may negatively impact the fusion of the scores when applied to held-out data.
The application of LR in this way can be seen as computing a trained weighted average of the two classifier scores.
Preliminary experiments evaluating max/min score selection of the clinical and LUS models, thereby optimising sensitivity or specificity respectively, consistently underperformed the strategies presented here and are not described further.
Additionally, preliminary experiments showed that model calibration also did not lead to any improvement in development set performance.

During fusion, all hyperparameters for the LUS classifier are held constant except the number of training epochs, the hidden dimensionality $h_d$, and whether the feature extractor is trained or frozen (reported in Table~\ref{tab:hyperparameters}).
For all fusion strategies, the effect of composing the clinical data feature vector from availability classes A, B, AB, or all features (ABC) before and after feature selection is considered.

\section{Results}

Firstly, the performance of the baseline ResNet LUS classifier (Section~\ref{subsec:lus_classifier}) is reported.
Next, Section~\ref{subsec:clinical_classifier} presents the performance of classifiers using only the clinical and demographic data. 
Finally, Section~\ref{subsec:fused_classifier} reports the performance after fusion of the two classifiers.

\subsection{LUS classifier}
\label{subsec:lus_classifier}

A grid-style hyperparameter optimisation on the development sets yields the values listed in Table~\ref{tab:hyperparameters}.
The development and independent test AUROCs for the resulting model (Baseline LUS) are presented alongside the clinical classifier in Table~\ref{tab:clinical_classifier_results}.
A test set AUROC of 0.91 [0.86, 0.96] is achieved by the LUS-only model, which is comparable to the value reported in previous work~\cite{ultrai2025escmid}.
We note again that, in the results reported here, clinically diagnosed TB cases are counted as TB positive, while in previous work they were not.
This performance serves as reference for the clinical and fusion classifiers.

\subsection{Clinical classifier}
\label{subsec:clinical_classifier}
The classification results achieved when training either LR or the MLP on the availability classes defined in Table~\ref{tab:clinical_features} are presented in Table~\ref{tab:clinical_classifier_results}.

Interestingly, classification using only self-reported values (A) outperforms classification based on measured values (B).
This may be due to the critical symptoms captured in class A, such as fever, cough, dyspnoea, weight loss, and known HIV, which are all captured in the features identified to improve development set performance during feature selection.

In Table~\ref{tab:clinical_classifier_results}, it is also clear that classes A and B provide mutually beneficial information.
This is evident because the union of all self-reported and measured values (development AUROC for AB after SFS is 0.87\ [0.83,0.90] for MLP) 
improves performance over training on either A or B alone (A: 0.82\ [0.78,0.86], or B:0.82\ [0.78,0.86]).

Leveraging all available collected values (ABC) and conducting feature selection results in the best overall development (0.89 [0.85,0.92]) and test-set performance (0.89 [0.83,0.96]), suggesting that although the POC tests considered here do not show great TB discrimination on their own (dev C: 0.57 [0.53, 0.61] for MLP) the information remains mutually beneficial relative to classes A and B.
It is important to note, however, that if the efficiency of the screening system is paramount, and the conduction of finger-prick based POC tests presents a barrier to screening, then a system which achieves only marginally worse performance using only self-reported and easily measurable values (AB) can still be trained.

\begin{table}[]
    \centering
    \caption{Development (95\% bootstrap CI~\cite{efron1994introduction}) and test set (95\% CI DeLong~\cite{delong_1988}) results for baseline LUS classifier and the various clinical classifiers trained on subsets of availability classes. A: self-reported, B: simple measurement, C: POC test.}
    \label{tab:clinical_classifier_results}
    \begin{tabular}{c c c c c c}
        \toprule
        Feature & Model & SFS & \# & Dev AUROC & Test AUROC \\
        subset            &&& Var \\
        \midrule
        \multicolumn{4}{c}{Baseline LUS} & 0.87 [0.84,0.90] & 0.91 [0.86,0.96]\\
        \midrule
        \multirow{4}{*}{A}  & LR    &                   & 28& 0.77    [0.72, 0.82]  & 0.86    [0.78, 0.93] \\
                            & MLP   &                   & 28& 0.78    [0.74, 0.82]  & 0.84    [0.77, 0.92] \\
                            & LR    & $\checkmark$      & 10& 0.82    [0.77, 0.85]  & 0.86    [0.79, 0.94] \\
                            & MLP   & $\checkmark$      & 13& 0.82    [0.78, 0.86]  & 0.80    [0.72, 0.89] \\
        \midrule
        \multirow{4}{*}{B}  & LR    &                   & 13& 0.81    [0.77, 0.86]  & 0.79    [0.70, 0.88] \\
                            & MLP   &                   & 13& 0.82    [0.78, 0.86]  & 0.77    [0.68, 0.86] \\
                            & LR    & $\checkmark$      & 7 & 0.82    [0.78, 0.86]  & 0.79    [0.70, 0.88] \\
                            & MLP   & $\checkmark$      & 5 & 0.82    [0.78, 0.86]  & 0.80    [0.72, 0.89] \\
        \midrule
        \multirow{4}{*}{C}  & LR    &                   & 3 & 0.53    [0.48, 0.57]  & 0.54    [0.45, 0.64] \\
                            & MLP   &                   & 3 & 0.57    [0.53, 0.62]  & 0.52    [0.42, 0.61] \\
                            & LR    & $\checkmark$      & 2 & 0.56    [0.52, 0.60]  & 0.54    [0.46, 0.61] \\
                            & MLP   & $\checkmark$      & 3 & 0.57    [0.53, 0.61]  & 0.52    [0.42, 0.61] \\
        \midrule
        \multirow{4}{*}{AB} & LR    &                   & 41& 0.83    [0.79, 0.87]  & 0.87    [0.80, 0.95] \\
                            & MLP   &                   & 41& 0.83    [0.79, 0.87]  & 0.86    [0.78, 0.94] \\
                            & LR    & $\checkmark$      & 19& 0.87    [0.84, 0.91]  & 0.87    [0.79, 0.95] \\
                            & MLP   & $\checkmark$      & 14& 0.87    [0.83, 0.90]  & 0.88    [0.81, 0.96] \\
        \midrule
        \multirow{4}{*}{AC} & LR    &                   & 31& 0.78    [0.74, 0.83]  & 0.88    [0.81, 0.95] \\
                            & MLP   &                   & 31& 0.79    [0.74, 0.83]  & 0.85    [0.78, 0.92] \\
                            & LR    & $\checkmark$      & 12& 0.83    [0.79, 0.87]  & 0.88    [0.82, 0.95] \\
                            & MLP   & $\checkmark$      & 15& 0.86    [0.82, 0.89]  & 0.83    [0.75, 0.91] \\
        \midrule
        \multirow{4}{*}{BC} & LR    &                   & 16& 0.81    [0.76, 0.85]  & 0.80    [0.71, 0.89] \\
                            & MLP   &                   & 16& 0.82    [0.78, 0.86]  & 0.81    [0.73, 0.89] \\
                            & LR    & $\checkmark$      & 9 & 0.82    [0.77, 0.86]  & 0.79    [0.71, 0.88] \\
                            & MLP   & $\checkmark$      & 10& 0.82    [0.78, 0.86]  & 0.80    [0.71, 0.89] \\
        \midrule
        \multirow{4}{*}{ABC}& LR    &                   & 44& 0.84    [0.79, 0.87]  & 0.88    [0.81, 0.95] \\
                            & MLP   &                   & 44& 0.84    [0.80, 0.88]  & 0.89    [0.82, 0.95] \\
                            & LR    & $\checkmark$      & 20& \textbf{0.89    [0.85, 0.92]}  & 0.89    [0.82, 0.96] \\
                            & MLP   & $\checkmark$      & 18& \textbf{0.89    [0.85, 0.92]}  & 0.89    [0.83, 0.96] \\
         \bottomrule
    \end{tabular}
\end{table}

On average over all the experiments presented in Table~\ref{tab:clinical_classifier_results}, the performance achieved by the MLP is marginally higher than that achieved by LR, yielding an average development AUROC of 0.79 compared to 0.78.
It is interesting that similar performance can be achieved using a model which is based only on a linear weighting of the input vector.
We nevertheless further consider the features selected by the MLP, both due to the slightly improved performance, and because the classification head of the fusion classifier is a two layer network, which is structurally more similar to the MLP we consider.

\subsubsection{Feature selection}
The best development AUROC is jointly achieved by LR and the MLP after SFS using all availability classes.
The 18 selected features are, in order of importance:
BMI, heart rate, hypertension, recent antibiotics, previous TB, weight loss, general state, days of cough, sex, malaria RDT, fever, night sweats, dyspnoea, HIV, pitting oedema, malaria RDT not done, age, and known HIV status.
Interestingly, two of the W4SS symptoms~\cite{world2021consolidated} (fever, night sweats) appear less informative for the clinical classifier than features such as heart rate, hypertension, antibiotics, which are not generally associated with TB screening.

When SFS is applied to the subset of features which are measured and self-reported (AB), the most to least informative were: BMI, heart rate, recent antibiotics, hypertension, previous TB, weight loss, general state, days of cough, sex, respiratory rate, night sweats, age, fever, and dyspnoea.
Except for respiratory rate, the features selected from the classes A and B are also selected from the pool of all features (ABC).
These selected features are all easy to capture, and reduce the number of necessary inputs by 27 values compared to a naive capture-all approach.

\subsection{Fused classifier}
\label{subsec:fused_classifier}

During fusion, only a subset of the hyperparameters is swept again, namely: whether the feature extractor is frozen, the hidden dimensionality $h_d$, and the number of training epochs.
The remaining hyperparameters are fixed to those which afforded the lowest average development set loss for the LUS-only classifier (Table~\ref{tab:hyperparameters}). 
This sweep is repeated independently for each availability class subset.
As the training of the LUS classifier is substantially more computationally intensive than the training of the clinical classifier, we do not repeat feature selection, but train the fused classifier either using the full feature vector or the subset of features identified during SFS with the MLP. 
Additionally, we do not further consider in isolation the subset of variables which require a POC test (C) or the combination of the self-reported/measured values and the POC testing class (AC,BC) as these were outperformed on the development sets by the more easily obtained classes A and B.

The results for all fusion experiments are presented in Table~\ref{tab:fusion_results}.
Overall, the inclusion of the clinical variables appears to improve average development set AUROC compared to the LUS only model, irrespective of the subset of features being used.
An absolute development set improvement of 1.03\% is achieved when using the concatenation strategy.
This rises to 2.09\% for the simple averaging strategy.

On average, over all the availability classes (before and after feature selection), the concatenation of the feature vector into the LUS classifier provides the smallest improvement of the three considered variants.
Interestingly, computing a simple average of the outputs of the two classifiers outperforms the concatenation strategy by 1.06\% absolute on average over all configurations on the development set.
Additionally, training an LR model to compute a weighted mean yields no discernible difference in development set performance, further making the average an attractive and simple alternative.

\begin{figure}
    \centering
    \includegraphics[width=\linewidth]{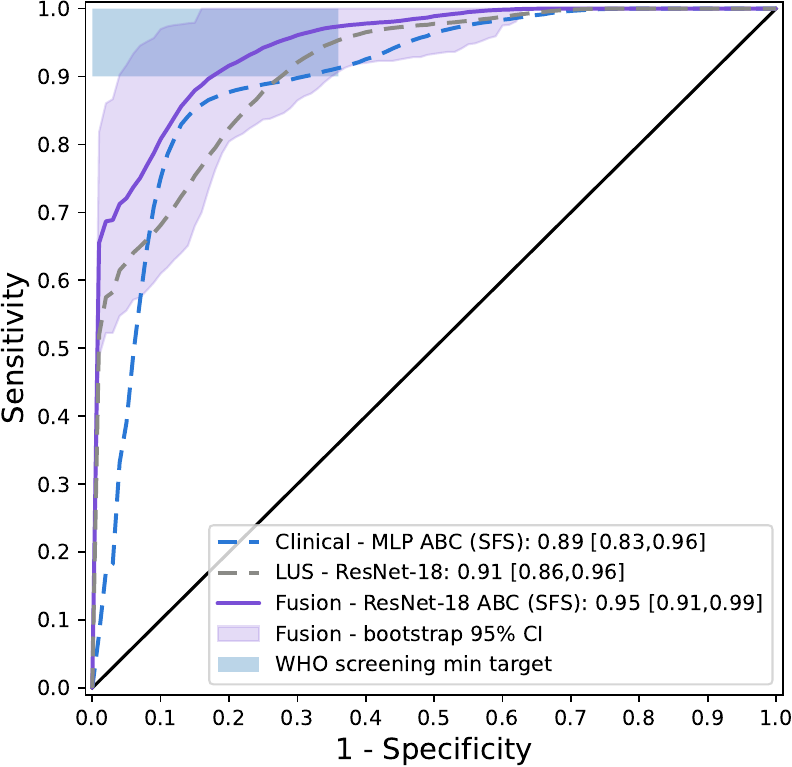}
    \caption{Test set receiver-operating characteristic (ROC) curve for the best of each considered architecture based on the development sets. All curves are smoothed over 10,000 bootstrap resamples and a bootstrap percentile-based 95\% CI is displayed for the fused classifier. AUROCs are presented with DeLong 95\% CI. Fusion: Mean of output probabilities from best clinical classifier MLP ABC (SFS) and ResNet-18. All three classifiers are those which afforded the lowest average development set loss across the respective configurations.}
    \label{fig:roc}
\end{figure}

For the fused classifier, the best overall development performance is again achieved by the classifier which incorporated the features selected from the pool of all variables (ABC).
This model achieves a test set AUROC of 0.95 [0.91,0.99], representing a 4\% absolute improvement over the baseline LUS classifier, although due to the size of the dataset statistical significance cannot yet be concluded (p=0.10).
For this classifier, the point estimate of the ROC curve, as presented in Figure~\ref{fig:roc}, appears to satisfy the WHO minimum specifications for a high sensitivity screening tool that forms part of a two step screening system~\cite{who_requirement}.
In fact, at 90\% sensitivity, the bootstrap 95\% confidence interval of the specificity falls completely within this range.
Furthermore, if the decision threshold is calculated as the average threshold which afforded 90\% sensitivity across the development sets, the corresponding test set sensitivity and specificity are 0.93 [0.81,0.98] and 0.76 [0.63,0.85] respectively, and 95\% confidence intervals are generated using Wilson's method~\cite{Wilson01061927}.
These point estimates are again within both the WHO minimum triage (0.7 specificity at 0.9 sensitivity) and screening (0.6 specificity at 0.9 sensitivity) guidelines~\cite{world2021consolidated,who_requirement}.

\begin{table}[]
    \centering
    \caption{Development (95\% bootstrap CI) and test set (95\% CI DeLong) results for baseline LUS classifier and the various fused classifiers trained on subsets of availability classes. A: self-reported, B: some simple measurement, C: POC test.}
    \label{tab:fusion_results}
    \begin{tabular}{c c c c c}
        \toprule
Feature subset & SFS & Fusion type & Dev AUROC & Test AUROC \\
\midrule
\multicolumn{3}{c}{Baseline LUS} & 0.87 [0.84,0.90] & 0.91 [0.86,0.96] \\
\midrule
\multirow{6}{*}{A}      &              & Concat & 0.87 [0.83,0.91] & 0.91 [0.85,0.97]\\
                        &              & LR     & 0.88 [0.84,0.91] & 0.94 [0.89,0.98]\\
                        &              & Mean   & 0.87 [0.83,0.90] & 0.93 [0.88,0.98]\\
                        & $\checkmark$ & Concat & 0.88 [0.85,0.91] & 0.92 [0.86,0.97]\\
                        & $\checkmark$ & LR     & 0.89 [0.86,0.92] & 0.93 [0.87,0.98]\\
                        & $\checkmark$ & Mean   & 0.89 [0.86,0.92] & 0.92 [0.87,0.98]\\
\midrule
\multirow{6}{*}{B}      &              & Concat & 0.89 [0.85,0.92] & 0.91 [0.86,0.96]\\
                        &              & LR     & 0.89 [0.85,0.92] & 0.90 [0.84,0.96]\\
                        &              & Mean   & 0.89 [0.86,0.92] & 0.89 [0.83,0.95]\\
                        & $\checkmark$ & Concat & 0.88 [0.85,0.91] & 0.90 [0.85,0.96]\\
                        & $\checkmark$ & LR     & 0.89 [0.86,0.92] & 0.91 [0.85,0.96]\\
                        & $\checkmark$ & Mean   & 0.89 [0.86,0.93] & 0.90 [0.84,0.96]\\
\midrule
\multirow{6}{*}{AB}     &              & Concat & 0.88 [0.84,0.91] & 0.94 [0.89,0.98]\\
                        &              & LR     & 0.89 [0.86,0.92] & 0.93 [0.88,0.98]\\
                        &              & Mean   & 0.89 [0.85,0.92] & 0.93 [0.88,0.98]\\
                        & $\checkmark$ & Concat & 0.88 [0.85,0.91] & 0.92 [0.86,0.98]\\
                        & $\checkmark$ & LR     & 0.90 [0.87,0.93] & 0.93 [0.88,0.98]\\
                        & $\checkmark$ & Mean   & 0.90 [0.87,0.93] & 0.94 [0.90,0.99]\\
\midrule
\multirow{6}{*}{ABC}    &              & Concat & 0.89 [0.85,0.92] & 0.93 [0.87,0.98]\\
                        &              & LR     & 0.90 [0.86,0.92] & 0.94 [0.90,0.98]\\
                        &              & Mean   & 0.89 [0.86,0.93] & 0.94 [0.89,0.98]\\
                        & $\checkmark$ & Concat & 0.90 [0.86,0.93] & 0.93 [0.88,0.99]\\
                        & $\checkmark$ & LR     & \textbf{0.91 [0.88,0.94]} & 0.94 [0.90,0.98]\\
                        & $\checkmark$ & Mean   & \textbf{0.91 [0.89,0.94]} & 0.95 [0.91,0.99]\\
\bottomrule
    \end{tabular}
\end{table}

\section{Conclusions}

In this work we have considered the effectiveness of incorporating clinical and demographic variables routinely collected during screening into a machine learning-based lung ultrasound (LUS) TB classification system.
Classifiers which leverage only clinical and demographic values which require some basic measurement or are self-reported yield test set AUROCs of 0.88 [0.81,0.96].
No substantial improvement is seen when adding measurements by POC tests (0.89 [0.83,0.96] p=0.61), although this additional set is in our experiments limited to only tests for HIV and malaria.
Feature selection is utilised to identify a subset of the most informative variables, 
thereby reducing the number which must ultimately be captured, and is therefore practically useful.
A simple average-based fusion of the output scores produced by a classifier for the clinical and demographic data and a second classifier for the LUS images is shown to consistently provide improved performance, increasing the test-set AUROC to 0.95 [0.91,0.99].
The average performance of this fused system falls within the WHO minimum screening (0.6 specificity at 0.9 sensitivity) guidelines achieving a specificity of 0.76 [0.63,0.85] and a sensitivity of 0.93 [0.81,0.98].

\section{Ethics and Data Availability}

The study that recruited the target cohort was approved by the committee for biomedical research ethics of the University of Parakou on 18 May 2021, under reference 0407/CLERB-UP/P/SP/R/SA, and is registered as NCT05423847 \cite{ultrai2025escmid}.
Every participant gave written informed consent, and a trusted representative assisted any participant who was unable to read or write.
Access to data is granted for research use under a data use agreement~\cite{ai-pocus_community_of_practice_2026}.

\section{Acknowledgements}

This project has received funding from the European Union’s Horizon Europe research and innovation Programme Global Health EDCTP3 Joint Undertaking under grant agreement No. 101190662 -- CAD LUS4TB, and has received funding from the Swiss State Secretariat for Education, Research and Innovation (SERI) under contract number 589188.
The TrUST study, within which the target cohort was recruited, is funded by the Swiss Lung Association.
We gratefully acknowledge financial support from Telkom (South Africa) for the research presented in this paper.
Computations were performed using the University of Stellenbosch's HPC1 (Rhasatsha).

The Butterfly iQ+ ultrasound equipment used in this project was provided by BFLY Operations, Inc.
BFLY Operations, Inc. had no involvement in the funding, data collection, design, analysis and outcome related to this project.

\section*{ORCID}

\small
\newcommand{\orcidline}[2]{%
  \textit{#1} & \href{https://orcid.org/#2}{%
    \includegraphics[width=1em]{orcid.pdf} orcid.org/#2}\\}
\noindent\begin{tabular*}{\linewidth}{@{}l@{\extracolsep{\fill}}r@{}}
\orcidline{Joshua M. Jansen van Vüren}{0000-0002-3406-4788}
\orcidline{Christiaan M. Geldenhuys}{0000-0003-0691-0235}
\orcidline{Devendra S. Parihar}{0000-0001-6404-6428}
\orcidline{Véronique Suttels}{0000-0002-5261-0439}
\orcidline{Trevor Brokowski}{0000-0002-7547-7749}
\orcidline{Ablo P. Wachinou}{0000-0002-1024-9665}
\orcidline{Mary-Anne Hartley}{0000-0002-8826-3870}
\orcidline{Rensu P. Theart}{0000-0003-0508-1690}
\orcidline{Grant Theron}{0000-0002-9216-2415}
\orcidline{Thomas R. Niesler}{0000-0002-7341-1017}
\end{tabular*}

\bibliography{references}

\end{document}